\documentclass[a4paper]{article}

\usepackage{ISCSLP2022}

\usepackage{amsmath,graphicx}
\usepackage{amssymb}
\usepackage{booktabs}
\usepackage{bm,amsfonts}
\usepackage{threeparttable}
\usepackage{float}
\usepackage{colortbl}
 \usepackage{enumitem}
\def\x{{\mathbf x}}

\def\x{{\bm x}}

\def\bpi{{\bm \pi}}
\def\B{{\mathcal B}}
\def\bt{{\bm \theta}}

\usepackage{hyperref}
\title{Exploiting Single-Channel Speech for Multi-Channel End-to-End Speech Recognition: A Comparative Study}

\name{Keyu An$^{1}$, Ji Xiao$^{2}$, Zhijian Ou$^{*, 1}$\thanks{$^{*}$ Corresponding author (ozj@tsinghua.edu.cn). This work is supported by NSFC 61976122.}}
\address{$^{1}$Speech Processing and Machine Intelligence (SPMI) Lab, Tsinghua University, Beijing, China\\
$^{2}$TasiTech Co., Ltd., Tianjin, China}

\begin{document}
\ninept
\maketitle
\begin{abstract}
Recently, the end-to-end training approach for multi-channel ASR has shown its effectiveness, which usually consists of a beamforming front-end and a recognition back-end. However, the end-to-end training becomes more difficult due to the integration of multiple modules, particularly considering that multi-channel speech data recorded in real environments are limited in size. This raises the demand to exploit the single-channel data for multi-channel end-to-end ASR. In this paper, we systematically compare the performance of three schemes to exploit external single-channel data for multi-channel end-to-end ASR, namely back-end pre-training, data scheduling, and data simulation, under different settings such as the sizes of the single-channel data and the choices of the front-end. Extensive experiments on CHiME-4 and AISHELL-4 datasets demonstrate that while all three methods improve the multi-channel end-to-end speech recognition performance, data simulation outperforms the other two, at the cost of longer training time. Data scheduling outperforms back-end pre-training marginally but nearly consistently, presumably because in the pre-training stage, the back-end tends to overfit on the single-channel data, especially when the single-channel data size is small.
\end{abstract}

\noindent\textbf{Index Terms}: multi-channel end-to-end ASR
%
\vspace{-0.3em}
\section{Introduction}
\label{sec:intro}
In recent years, significant progress has been made in automatic speech recognition (ASR). However, speech recognition in far-field scenarios is still a challenging task~\cite{chime4,chime6}. To be specific, Fu~\emph{et al.}~\cite{aishell4} report an over 30\% character error rate (CER) on AISHELL-4, which is recorded in distant-talking conference scenario, and Watanabe~\emph{et al.}~\cite{chime6}  report an over 50\% word error rate (WER) on CHiME-6, which is recorded in everyday home environments with distant microphones. 

Leveraging multi-channel signals with beamforming has been shown to improve speech recognition performance in far-field scenarios~\cite{chime4,chime6,Beamnet_heymann,Multich-e2e}. Conventionally, the beamforming front-end and the ASR back-end are optimized separately under different criteria. The enhanced output produced by the beamformer is then processed with the single-channel ASR back-end. This conventional approach is known to have the drawback that the optimization objectives for the two sub-tasks are not matched and the information flow between them is unduly limited \cite{seltzer2004likelihood,zhao2007closely}.
Recently, this drawback has been re-visited by building multi-channel speech recognition system with a unified neural network~\cite{Beamnet_heymann,Multich-e2e,joint_tencent,3dcnn}. In such models, gradients from the back-end can be propagated to the front-end, which is typically a neural beamformer. 

Despite promising results, training a multi-channel end-to-end ASR system, which is composed of several modules, has been shown to be much more difficult than training a single-channel ASR system~\cite{Beamnet_heymann,mimo,branches}. Specifically, straightforward optimization of a multi-channel end-to-end system often does not converge~\cite{unified}, or leads to sub-optimal results~\cite{Beamnet_heymann}. Another difficulty is that multi-channel data are relatively more expensive to collect. To address these issues, exploiting single-channel data in multi-channel end-to-end systems has been studied. For example, it is suggested that the front-end and the back-end are initialized with the corresponding \textbf{pre-trained} models respectively~\cite{Beamnet_heymann,unified}. In MIMO-SPEECH~\cite{mimo}, \textbf{data scheduling}, which means every batch is randomly chosen either from the multi-channel set or from the single-channel set during training, is proposed to regularize the training process. It has been shown that the multi-channel ASR can benefit from training on the \textbf{simulated multi-channel data}~\cite{chime4}.
Although the three methods (pre-training, data scheduling and data simulation) of using external single-channel data in multi-channel ASR systems are previously known in the literature,  previous studies have not yet evaluated and compared them systematically.
The results from previous individual works are not directly comparable to each other, since they are not evaluated in a common experimental setup.
On top of these previous studies, the main contributions of this paper are: 

1) We conduct a suite of experiments to compare the three methods in utilizing single-channel speech to improve the multi-channel end-to-end ASR system. Thorough analyses under different experiment settings are given, including the choices of the front-end and the amounts of the single-channel speech data. It is found that data scheduling outperforms back-end pre-training marginally but nearly consistently, presumably because that in the pre-training stage, the back-end tends to overfit on the single-channel data. Data simulation is more expensive but outperforms the other two, which may be due to the fact that the simulated multi-channel data augment the training data for both front-end and back-end, while back-end pre-training and data scheduling only augment the training data for back-end.

2) We integrate the multi-channel speech enhancement front-end with the CTC-CRF\footnote{Other single-channel speech recognition models such as attention based encoder-decoder (AED) and RNN-transducer could be used. The comparison under these frameworks are left for further study.} based single-channel speech recognition back-end \cite{CRF_IC19,CAT}, and develop a multi-channel end-to-end speech recognition system. 
With data augmentation methods such as SpecAugment~\cite{specaug} and WavAugment~\cite{wavaug}, the resulting system obtains competitive results on CHiME-4 and AISHELL-4 benchmarks.

The rest of the paper is organized as follows. 
Section 2 outlines related work. 
Section 3 describes our multi-channel end-to-end speech recognition system.
The three methods to exploit single-channel data are detailed in Section 4.
Section 5 introduces the experiment settings on the CHiME-4 and AISHELL-4 tasks, and the results are shown in Section 6. 
Section 7 presents the conclusion. 

\section{Related Work} 
\label{sec:related-works}
Recently, there have been growing interests in building end-to-end systems for multi-channel ASR, which can be categorized into two approaches: the approach of multi-channel acoustic modeling without an explicit beamformer and the neural beamformer based approach. In multi-channel acoustic modeling without an explicit beamformer, the neural network is viewed as a replacement for the conventional beamformer. For example, ~\cite{concatenate} directly concatenates the multi-channel features and improves far-field ASR performance over single-channel input. However, such direct concatenation of the multi-channel features burdens the training of the neural network with increasing number of parameters~\cite{concatenate} to capture the complex relationships between microphones. Several novel neural operations were proposed to alleviate this issue, such as 3-D CNN~\cite{3dcnn} and quaternion neural networks~\cite{qnn}. In general, these methods require complex design of model architectures, and the models depend on the microphone configurations. Therefore, once the number and order of microphone channels are changed, the neural network has to be reconfigured and retrained.

In neural beamformer based approach,  the neural beamformer is cast as a differentiable component to allow joint optimization of the multi-channel speech enhancement with the ASR criterion. This approach can be further categorized into two types. The first type is the  mask estimation method. The neural network is used to estimate the time-frequency masks, which are used to compute the statistics of speech and noise. Using these statistics, the filter coefficients are then computed within the framework of constrained optimization for noise reduction, such as minimum variance distortionless response (MVDR) and generalized eigenvalue (GEV) ~\cite{Beamnet_heymann,Multich-e2e}. 
The second type is the filter estimation method. The neural network is used to estimate the filter coefficients directly~\cite{deep_beamforming_xiao,adaptive_beamforming}. In the filter estimation method, the filter coefficients are less restricted than in the mask estimation method, and their estimation becomes more difficult due to the high freedom of the filter coefficients~\cite{Multich-e2e,deep_beamforming_xiao}. Moreover, similar to the multi-channel acoustic model without an explicit beamformer, the filter estimation network also depends on the microphone configurations. In light of the above trends,  we choose the mask based neural beamformer to build the multi-channel end-to-end speech recognition system in this comparative study.

\section{Multi-Channel End-to-End Speech Recognition}
In this section, we introduce our multi-channel end-to-end speech recognition system, including the mask-based MVDR neural beamformer as the front-end (FE), the CTC-CRF based AM as the back-end (BE), and the overall processing pipeline. 
\subsection{The MVDR based front-end} \label{sec:mvdr}
We adopt the state-of-the-art MVDR neural beamformer ~\cite{Multich-e2e,MVDR} as the front-end. MVDR reduces the noise and recovers the signal component by applying a linear filter to the overall observation vector:
\begin{equation} \label{eq:beforming} 
\hat{x}(t,f) = \sum^{C}_{c=1} h(f,c) \times x(t,f,c)
\end{equation}
where $ x(t,f,c)\in \mathbb{C} $  denotes the short-time
Fourier transform (STFT) coefficient at time-frequency bin $(t, f)$ of the noisy signal at microphone $c$.  $ \hat{x}(t,f)  \in \mathbb{C} $
is the enhanced STFT coefficient, and $C$ is the numbers of microphones. According to the MVDR formulation ~\cite{MVDR},  the time-invariant filter coefficient ${\rm \textbf h}(f) = \{h(f,c)\}_{c=1}^C \in \mathbb{C}^C $ is obtained by
$$
{\rm \textbf h}(f) = \frac{{{\bf \Phi_{NN}^{-1}}(f)}{ \bf \Phi_{SS}}(f)} {{\rm tr} \{ {{\bf \Phi_{NN}^{-1}} (f)}{ \bf \Phi_{SS}}(f)\}} { \rm \textbf u} 
$$
Here ${\bf \Phi_{SS}} (f) \in \mathbb{C}^{C \times C} $ and ${\bf \Phi_{NN}} (f) \in \mathbb{C}^{C \times C} $ are the cross-channel power spectral density (PSD) matrices (also known as spatial covariance matrices) for speech and noise signals respectively, which are estimated via the mask-based approach~\cite{mask_heymann}. ${ \rm \textbf u} \in \{0,1\}^C $ is the one-hot vector,  indexing the reference microphone, which can be selected by principal component analysis ~\cite{mask_heymann} or neural networks ~\cite{Multich-e2e}. $ \rm tr\{  \cdot  \} $  denotes matrix trace.

\subsection{The CTC-CRF based back-end}
For the acoustic model (AM) for recognition, we adopt the newly developed CTC-CRF~\cite{CRF_IC19}. Like CTC, CTC-CRF defines the posterior of the label sequence $\bm{l} $ as the sum of the posterior of the hidden state sequence $\bpi$ as follows:
\begin{equation} \label{eq:post-l} 
	p_{\bt}(\bm{l} | \x) = \sum_{\bpi \in \mathcal{B}^{-1}(\bm{l} )} p_{\bt}(\bpi | \x)
\end{equation}
where $ \B$ denotes the mapping that removes consecutive repetitive labels and blanks in the hidden state sequence.

The posterior of $\bpi$ is further defined by a CRF:
\begin{equation} \label{eq:post-pi}
p_{\bt}(\bpi|\x) = \frac{\exp(\phi_{\bt}(\bpi, \x))}{\sum_{\bpi'}{\exp(\phi_{\bt}({\bpi', \x}))}}
\end{equation}
Here $\phi_{\bt}(\bpi, \x)$ denotes the potential function of the CRF, defined as follows:
\begin{equation} \label{eq:potential}
\phi_{\bt}(\bpi, \x) = \log p( \B(\bpi) )+ \sum_{t=1}^{T} \log p_{\bt}(\pi_t|\x)
\end{equation}
where $\sum_{t=1}^{T} \log p_{\bt}(\pi_t|\x)$ defines the node potential, calculated from the AM, 
and $\log p( \B(\bpi) )$ defines the edge potential, realized by an n-gram language model of labels. By incorporating $\log p( \B(\bpi) )$ into the CRF potential function, the undesirable conditional independence assumption in CTC is naturally avoided. It has been shown that CTC-CRF outperforms regular CTC consistently on a wide range of benchmarks, and is on par with other state-of-the-art end-to-end models \cite{CRF_IC19,CAT,CTC-CRF-NAS}.
\subsection{The multi-channel end-to-end model}
\label{sec:joint}
On top of the MVDR based FE and the CTC-CRF based BE, we build a unified architecture for multi-channel end-to-end speech recognition, and apply joint optimization (JO) of front-end and back-end. Thus, the training loss is defined as:
\begin{equation} \label{eq:e2e-obj}
\mathcal{L}(\bt) = - \log p_{\bt}(\bm{l}| {\rm Feature} (\hat{\x}))
\end{equation}
where $\hat{\x}$ is obtained by Eq. (\ref{eq:beforming}), and $p_{\bt}$ is defined by combining Eq. (\ref{eq:post-l}) $\sim$ (\ref{eq:potential}). Feature(·) is the feature extraction and pre-processing function, including log fbank transformation, normalization, and subsampling, as detailed in Section~\ref{setting}. 

\section{Methods to exploit single-channel Speech}
\subsection{Back-end pre-training}
\label{sec:pre}
The first approach to exploiting single-channel speech data in multi-channel end-to-end speech recognition system is to do back-end pre-training~\cite{Beamnet_heymann}. Specifically, the training process consists of two stages. In the first stage, the back-end AM is trained with single-channel speech data. In the second stage, we perform joint optimization described in Section ~\ref{sec:joint}. In the multi-channel end-to-end speech recognition model, the enhanced log fbank feature produced by the front-end is supposed to be similar to the log fbank of single-channel (clean) speech~\cite{mimo}. Thus, back-end pre-training is expected to provide a better initialization for the AM, compared with random initialization. However, the back-end may overfit on the single-channel (clean) speech and degrade the performance of joint optimization, especially when pre-training is performed on a small single-channel dataset, as shown later in our experiments.
\subsection{Data scheduling}
The second approach is data scheduling~\cite{mimo}. Different from the two-stage pipeline described in Section~\ref{sec:pre}, training with data scheduling is conducted in a single stage.  In data scheduling, the training data comes from two sources: the multi-channel set and the single-channel set. When the training batch comes from the multi-channel set, we perform joint optimization described in Section ~\ref{sec:joint}. Otherwise, we bypass the front-end and only optimize the single-channel AM over the single-channel batch. Under this mechanism, the back-end AM is trained with both the original (unenhanced) and the enhanced single-channel data, and would be more robust to the input variations \cite{mimo}. Moreover, the back-end can be independently optimized when the front-end is bypassed, which regularizes the training process ~\cite{mimo} and eases the training. 
In practice, we set the ratio between the single-channel batch size and the multi-channel batch size to be $\rm \frac{\#utts~of~single-channel~data}{\#utts~of~multi-channel~data}$, so that we can sweep over the whole single-channel data and the whole multi-channel data with equal number of batches. 
\subsection{Data simulation}
\label{sec:simu}
The third approach is to simulate multi-channel data using single-channel data~\cite{pyroomacoustics}. In this approach, we first define a room to which the sound source (the single-channel wave samples) and a microphone array are attached. Then, a simulation method is used to create artificial room impulse responses (RIRs) between the source and microphones. The microphone signals are then created by convolving the single-channel wave samples with the RIRs. After data simulation, the simulated multi-channel data are mixed with the real multi-channel data to train the multi-channel speech recognition system. 
Different from back-end pre-training and data scheduling, where the single-channel data is used only for training the back-end, using simulated multi-channel data can augment the training data for the front-end, but at the cost of much longer training time. This increasing time cost is due to 1) the simulation time, 2) the additional time costs for front-end computation, online feature extraction and processing, as explained in \ref{sec:joint}. We compare the time costs for the three methods in Table~\ref{summ}.
\begin{table}
	\centering
	\caption{Comparison of the time costs for back-end pre-training (PT), data scheduling (DS), and data simulation (Simu). Single-stage denotes single-stage training, and augment FE indicates whether the method augments the training data to feed to the front-end (FE). $T_1$ is the time cost for multi-channel end-to-end joint optimization per epoch, $T_2$ is the time cost for training back-end using single-channel data per epoch, and $N = \rm \frac{{\#}utts~of~single-channel~data}{ {\#}utts~of~multi-channel~data}$. $T_1$ $\gg$ $T_2$ as explained in Section~\ref{sec:simu}. For exapmle, in AISHELL-4 experiments with AISHELL-1 as single-channle data, $T_1 \
\approx 10 T_2$, $ N \approx 1.2 $. Note that the time for pre-training back-end in PT and simulating data in Simu are not shown in the table.} 
	\vspace{-0.25cm}
	\scalebox{0.9}{
	\begin{tabular}{cccc}
		\toprule
        \textbf{methods} & \textbf{single-stage}  & \textbf{aug FE}  & \textbf{time cost per epoch}\\
        \midrule
		PT & No &  No & $T_1$\\
		DS & Yes & No &  $T_1 + T_2$\\
		Simu &  Yes & Yes & $(1 + N)T_1$ \\
		\bottomrule
	\end{tabular}}
	\label{summ}
	\vspace{-0.2cm}
\end{table}

\begin{table}
	\centering
	\caption{Effect of joint optimization (JO) of front-end and back-end, measured by word error rates (WERs) on CHiME-4. Further improvement can be obtained by data augmentation in the JO framework.
	Baseline denotes the baseline system provided by CHiME-4 challenge. }
	\vspace{-0.25cm}
	\scalebox{0.8}{
	\begin{tabular}{cccccc}
		\toprule
		\textbf{FE}   &  \textbf{JO} & \textbf{Dev real} & \textbf{Dev simu} & \textbf{Eval real} & \textbf{Eval simu}\\
		\midrule
		Baseline~\cite{chime4} & No & 8.14 &	9.07 & 15.00 & 14.23 \\
        \midrule
		BeamformIt  & No & 7.28 & 7.98 & 11.11 & 11.97 \\
		MVDR  & No & 6.95 & 8.08 & 10.50 & 11.03 \\
		\midrule
		MVDR   & Yes & 6.15 & 5.61 & 9.29 & 6.14 \\
		\multicolumn{2}{c}{\quad + SpecAug} & 5.93 & 5.04 & 8.42 & 6.00 \\
		\multicolumn{2}{c}{\quad  \qquad + WavAug} & 5.60 & 4.94 & 8.06 & 5.70 \\
		\bottomrule
	\end{tabular}}
	\label{joint optimization}
	\vspace{-0.5cm}
\end{table}

\begin{table*}
\caption{Comparison of back-end pre-training (PT), data scheduling (DS) and data simulation (Simu) on CHiME-4, with joint optimization (JO) applied or not. Average denotes the WERs averaged on the four test sets Dev/Eval-real/simu.}
\vspace{-0.25cm}
\begin{threeparttable}
	\centering
	\scalebox{0.98}{
	\begin{tabular}{ccccccccc}
		\toprule
        \textbf{ID} & \textbf{FE pre-training data} & \textbf{Single-channel data}&  \textbf{method} & \textbf{Dev real} & \textbf{Dev simu} & \textbf{Eval real} & \textbf{Eval simu} &\textbf{Average}\\
        \midrule
		0 & No pre-training & None  & JO & 5.60 & 4.94 & 8.06 & 5.70 & 6.08 \\
		\midrule
        1 & No pre-training & WSJ & PT + JO & 6.00 & 5.41 & 8.37 & 6.12 & 6.48 \\
        2 & No pre-training & WSJ & DS + JO & 4.95 & 4.60 & 7.38 & 5.55 & 5.62\\     
        3  & No pre-training & WSJ & Simu + JO & 5.21 & 4.41 & 7.10 & 4.82 & 5.39 \\
        4 & CHiME-4 & WSJ & PT & 15.93 & 23.24 & 31.80 & 29.79 & 25.19 \\
        5 & CHiME-4 & WSJ & PT + JO & 5.83 & 5.51 & 8.44 & 5.91 & 6.42\\
        6 & CHiME-4 & WSJ & DS + JO & 4.83 & 4.68 & 7.38 & 5.79 & 5.67\\

        \midrule
        7 & No pre-training & Librispeech & PT + JO& 4.81 & 4.83 & 7.66 & 5.48 & 5.70 \\
        8 & No pre-training & Librispeech & DS + JO & 4.35 & 4.43 & 6.10 & 4.80 & 4.92 \\
        9  & No pre-training & Librispeech & Simu + JO &  4.11 & 4.23  &  6.16 &  4.59 & 4.77 \\
        10 & CHiME-4 & Librispeech &  PT  & 4.28 & 4.57 & 5.13 & 6.71 & 5.17 \\
        11 & CHiME-4 & Librispeech &  PT + JO &  4.17 & 4.36 & 6.60 & 4.73 & 4.97\\
        12 & CHiME-4 & Librispeech & DS + JO & 4.24  & 4.34 &  6.22 & 4.71 & 4.88 \\

		\bottomrule
	\end{tabular}}
	\vspace{-0.25cm}
\end{threeparttable}
		\label{comp}
\end{table*}
\begin{table}
	\centering
	\caption{The character error rate (CER) results on AISHELL-4. The single-channel data is AISHELL-1. For all experiments, the front-end is not pre-trained.}
	\vspace{-0.25cm}
	\scalebox{1}{
	\begin{tabular}{cc}
		\toprule
        \textbf{method} & \textbf{eval CER} \\
        \midrule
		JO W/O single-channel data & 60.7 \\
		PT + JO & 37.3 \\
		DS + JO &  37.0 \\
		Simu + JO & 36.5 \\
		\bottomrule
	\end{tabular}}
	\label{tab:aishell4}
	\vspace{-0.5cm}
\end{table}
\section{Experiment settings}
\subsection{Datasets}
\setlist{leftmargin=3mm}
\subsubsection{Multi-channel Datasets}
\begin{itemize}
\setlength{\itemsep}{0pt}
\setlength{\parsep}{0pt}
\setlength{\parskip}{0pt}
\item \textbf{CHiME-4}~\cite{chime4}  is a speech recognition task in public noisy environments, recorded using a tablet with a 6-channel microphone array. The corpus is in English and the training data length is 18 hours.
\item \textbf{AISHELL-4}~\cite{aishell4} is a multi-channel mandarin dataset for conversation speech in conference scenarios, containing 118 hours of meeting recording, recorded using an 8-channel microphone array. As the integration of the speaker diarization module is beyond the scope of this paper, we select the non-overlapped part of the training ($\sim$ 50 hours) and evaluation set of AISHELL-4 according to the ground-truth segmentation information. For running the AISHELL-4 experiments, we use the open-source lexicon provided by AISHELL-1 dataset as there is no official lexicon in the AISHELL-4 dataset, and word segmentation over transcripts are performed using the Jieba segmentation toolkit \footnote{{https://github.com/fxsjy/jieba}}.
\end{itemize}
\subsubsection{Single-channel Datasets}
\begin{itemize}
\setlength{\itemsep}{0pt}
\setlength{\parsep}{0pt}
\setlength{\parskip}{0pt}
\item \textbf{WSJ}~\cite{wsj} contains about 80 hours of English training data recorded under clean conditions. 
\item \textbf{Librispeech}~\cite{librispeech} contains 1000 hours of English read speech, derived from audiobooks.
\item \textbf{AISHELL-1}~\cite{aishell} is a 178-hour mandarin speech corpus.
\end{itemize}
\subsection{Settings}
\label{setting}
We use the CTC-CRF based ASR Toolkit - CAT~\cite{CAT} to conduct the experiments. In our experiment, the inputs to the front-end are STFT features. The mask estimation network in the neural beamformer is a 3-layer BLSTM. After beamforming, the enhanced single-channel STFT features are firstly converted to 40-dimensional log fbank features, and then mean-variance normalized. The normalized log fbank features are appended with delta and delta-delta features and subsampled by a factor of 3. Similar to~\cite{CAT}, the acoustic model is two blocks of VGG layers followed by a 6-layer BLSTM, and the BLSTM has 320 hidden units per direction. Speed perturbation is adopted for data augmentation.

In data simulation experiments, we adopt pyroomacoustics~\cite{pyroomacoustics} to simulate multi-channel waves using single-channel speech as the source signal.  We define a 10m $\times$ 7.5m $\times$ 3.5m room, and the source is located at [2.5, 3.73, 1.76]. The microphone configurations are the same as the ones used in CHiME-4 challenge and AISHELL-4 challenge, respectively.

Note that we did not use the transcripts of single-channel data in language model building. In CHiME-4 experiments, we use the 3-gram language model provided by the challenge. In AISHELL-4 experiments, we use a 3-gram language model trained on the AISHELL-4 training set transcripts.

\section{Experiments}
\subsection{Joint optimization of front-end and back-end}
The effect of joint optimization (JO) of front-end and back-end over CHiME-4 is shown in Table~\ref{joint optimization}. 
Two different front-ends are tested in separate optimization, namely the delay-and-sum beamformer (BeamformIt~\cite{beamformit}) and the MVDR based neural beamformer (Sec. \ref{sec:mvdr}).
It can be seen that the jointly optimized model significantly outperforms the separately optimized models, which confirms the superiority of joint optimization.

In the JO framework, further improvement can be obtained by data augmentation.
We apply WavAugment~\cite{wavaug} on the input audio to the front-end, and SpecAugment~\cite{specaug} on the input log fbank features to the back-end. It can be seen from Table~\ref{joint optimization} that introducing WavAugment and SpecAugment in JO yields the best performance. 
In the following experiments, WavAugment and SpecAugment are adopted by default.

\subsection{Comparison of back-end pre-training, data scheduling and data simulation}
We compare back-end pre-training (PT), data scheduling (DS), and data simulation (Simu) on CHiME-4, using different single-channel data (WSJ or Librispeech) and different front-ends (pre-trained or random initialized\footnote{Pretrained FE means that training the mask estimation network using multi-channel speech and the corresponding clean speech. As can be seen from Table~\ref{comp}, the performances between pretrained and random initialized for FE are close to each other.}), as shown in Table~\ref{comp}. It can be seen that DS outperforms PT nearly consistently. Notably, PT on WSJ (Exp 1 and Exp 5) even degrades the performance, compared with the model jointly trained from scratch (Exp 0). This is probably because in the pre-training stage, the back-end tends to overfit on the single-channel WSJ data. Also note that pre-training on WSJ without joint optimization on CHiME4 can lead to quite poor performance (Exp 4).
When we increase the size of the single-channel data by using Librispeech, PT produces substantial improvement (Exp 7 and Exp 11). However, the results by PT are still inferior to DS in terms of average WERs. 
Data simulation produces superior results than DS and PT (Exp 3 and Exp 9), but at the cost of increased training time. Specifically, it takes about 114 hours for 1 epoch when training with multi-channel data simulated from Librispeech, while only 1.5 hours for PT + JO, and 3 hours for DS + JO, in our experiment. Similar conclusions can be drawn from the results on AISHELL-4 (Table \ref{tab:aishell4}).

\section{Conclusion}
In this paper, three methods to exploit single-channel data for multi-channel end-to-end ASR are systematically compared. It is found that data scheduling, which performs joint optimization using multi-channel data and back-end training using single-channel data in one stage, outperforms the two-stage pre-training plus joint optimization method, presumably because that in the back-end pre-training stage, the back-end tends to overfit on the single-channel (clean) data, especially when the single-channel data is limited in size. Data simulation outperforms the other two, but at the cost of longer training time. We hope these findings would be helpful for future work to further explore
better methods to efficiently leverage single-channel data for multi-channel speech recognition.
\bibliographystyle{IEEEtran}
\bibliography{mybib}

\end{document}